\documentclass[showpacs,preprintnumbers,amsmath,amssymb]{revtex4}

\usepackage{graphicx}
\usepackage{dcolumn}
\usepackage{bm}
 
\def\mbi#1{\mbox{\bfseries\itshape #1}} 

\begin{document}

\preprint{APS/123-QED}

\title{Constraints on the Primordial Magnetic Field from $\sigma_8$}

\author{Dai G. Yamazaki$^{1}$}
 \homepage{http://th.nao.ac.jp/~yamazaki/}
 \email{yamazaki@th.nao.ac.jp}
\author{Kiyotomo Ichiki$^{2}$}%
\author{Toshitaka Kajino$^{1,3}$}%
\author{Grant. J. Mathews$^{4}$}%
\affiliation{%
$^{1}$National Astronomical Observatory of Japan
2-21-1 Osawa, Mitaka, Tokyo 181-8588, Japan
}%
\affiliation{%
$^{2}$Research Center for the Early Universe,
School of Science, the University of Tokyo,
Hongo 7-3-1, Bunkyo-ku, Tokyo 113-0033, Japan
}%
\affiliation{%
$^{3}$
Department of Astronomy, School of Science, The University of Tokyo,
Hongo 7-3-1, Bunkyo-ku, Tokyo 113-0033, Japan
}%
\affiliation{%
$^{4}$Center for Astrophysics,
Department of Physics, University of Notre Dame, Notre Dame, IN 46556, U.S.A.
}%

\date{\today}

\begin{abstract}
A primordial magnetic field (PMF) can affect the evolution of density field fluctuations in the early universe.
In this paper we constrain the PMF amplitude $B_\lambda$ and power spectral index $n_\mathrm{B}$ by comparing calculated density field fluctuations with observational data, i.e.~the number density fluctuation of galaxies.
We show that the observational constraints on cosmological density fluctuations, as parameterized by  $\sigma_8$, lead to  strong constraints on the amplitude and spectral index of the PMF.
\end{abstract}

\maketitle
\section{Introduction}
Magnetic fields have been observed 
\cite{Kronberg:1992pp,Wolfe:1992ab, Clarke:2000bz,Xu:2005rb}
in clusters of galaxies  with a strength of $0.1-1.0~\mu$ G. 
One possible explanation for such 
magnetic fields in galactic clusters is the existence of a  primordial magnetic field (PMF) of order  1 nG  whose field lines collapse as structure forms. 
Therefore, recently, the origin and amplification mechanism of the PMF in the scale of galaxy cluster have been proposed and studied intensively by a number of authors
\cite{
Betschart:2003bn,
Boyanovsky:2002kq,
Bruni:2002xk,
Bamba:2004cu,
Berezhiani:2003ik,
Dolgov:2005ti,
Hanayama:2005hd,
Takahashi:2005nd,
ichiki:2006sc,
Quashnock:1989jm}. 
The origin and detection of the PMF is, hence, a subject of considerable interest in modern cosmology. 
Moreover, the PMF could influence a variety of phenomena in the early universe
\cite{Grasso:2000wj,Takahashi:2007ds,Ichiki:2007hu,Takahashi:2008gn} such as the cosmic microwave background (CMB)
\cite{
Yamazaki:2005yd,
Yamazaki:2004vq,
Yamazaki:2006bq,
Yamazaki:2006ah,
Yamazaki:2007oc,
Kojima:2008rf,
Mack:2001gc,
Subramanian:1998fn,
Subramanian:2002nh,
Lewis:2004ef,
Kosowsky:2004zh,
Kahniashvili:2005xe,
Kahniashvili:2006hy,
2008PhRvD..78f3012K,
Challinor:2005ye,
Dolgov:2005ti,
Gopal:2005sg,
Giovannini:2006kc,
Giovannini:2008aa}, 
 and  the formation of large-scale structure (LSS)
\cite{
Yamazaki:2006mi,
Tsagas:1999ft,
Giovannini:2004aw,
Tashiro:2005hc,
Sethi:2008eq}.

If dynamically significant large-scale magnetic fields were present in 
the early universe, they would have affected the formation and 
evolution of the observed structure.   Thus, some signatures of 
the existence of a PMF  should be apparent in the presently observed cosmic structure. 

In this regard, the alternative normalization parameter $\sigma_8$ is of particular interest.  It is defined \cite{Peebles:1980booka} as the root-mean-square of the matter density fluctuations in a comoving sphere of radius $8h^{-1}$ Mpc.  It is determined by a weighted integral of the matter power spectrum. 
Observations which determine $\sigma_8$ provide information about the physical processes affecting the evolution of density-field fluctuations  and the formation of structure on the cosmological scales. 
The mechanisms by which a PMF can affect the density field fluctuations on cosmological scales has been described in our previous work \cite{Yamazaki:2006mi}.  Of course, $\sigma_8$ is also affected by the presence of a PMF. 
In this article we show that by
considering the effect of a PMF on $\sigma_8$ and comparing theoretically estimated values for $\sigma_8$ with the  observed range, we can obtain not only insight into the underlying  physical processes of density field fluctuations in the presence of a PMF,  but also  place constraints on the amplitude and spectral index of the PMF. 

\section{The Model}
We use the isocurvature magnetized initial conditions with
adiabatic relations for the fluids evolution of primary density perturbations and in the presence of a PMF.
For the present purposes we fix the cosmological parameters to those of
the best-fit flat $\Lambda$CDM model as given in Ref.~\cite{Dunkley:2008ie}, i.e.~$h=0.719$, 
$\Omega_bh^2=0.02273$, $\Omega_c h^2=0.1099$, $n_S=0.963$,
 and $\tau_c=0.087$,
where $h$ denotes the Hubble parameter in units of 100 km s$^{-1}$Mpc$^{-1}$, $\Omega_b$
and $\Omega_c$ are the baryon and cold dark matter densities in units of
the critical density, $n_S$ is the spectral index of the primordial
scalar fluctuations, and $\tau_c$ is the optical depth for
Compton scattering. We use natural units $c=\hbar=1$.
\subsection{Primordial Magnetic Field}
Before recombination, Thomson scattering between photons and
electrons along with Coulomb interactions between electrons and
baryons were sufficiently rapid to ensure that the photon-baryon plasma
behaved as a single tightly coupled fluid.
Since the trajectories of plasma particles are bent by Lorentz forces in a
magnetic field, photons are indirectly influenced by the magnetic field
through Thomson scattering. 
The energy density of the magnetic field can be treated as a first order
perturbation upon a flat Friedmann-Robertson-Walker (FRW) background
metric.  
In this linear approximation\cite{Durrer:1999bk}, 
the magnetic field evolves as a stiff
source.
Therefore, we can discard all back reaction terms from the magnetohydrodynamic (MHD) fluid onto the field itself.
\subsection{Power Spectrum from the PMF}
During  the epochs of interest here,
the conductivity of the primordial plasma is very large, and the PMF is "frozen-in" to a very good approximation\cite{Mack:2001gc}. 
 Furthermore, we can neglect the electric field, i.e.~$E\sim 0$, and can 
 decouple the time evolution of the magnetic field from its spatial dependence, i.e. $\mathbf{B}(\mathbf{x},\tau) = \mathbf{B}(\mathbf{x})/a^2$ for very large scales. This leads to the following simplified electromagnetic energy-momentum tensor,
\begin{eqnarray}
{T^{00}}_{[\mathrm{EM}]}(\mathbf{x},\tau)=\frac{B(\mathbf{x})^2}{8\pi a^6} \label{eq_MST_00}~~, \\
{T^{i0}}_{[\mathrm{EM}]}(\mathbf{x},\tau)={T^{0k}}_{[\mathrm{EM}]}(\mathbf{x},\tau)=0 \label{eq_MST_0s} ~~,\\
-{T^{ik}}_{[\mathrm{EM}]}(\mathbf{x},\tau)=\sigma^{ik}_\mathrm{B}=
\frac{1}{8\pi a^6}\left\{
	2B^i(\mathbf{x}) B^k(\mathbf{x}) -
\delta^{ik}B(\mathbf{x})^2
\right\}~~.
\label{eq_MST_ss}
\end{eqnarray}
We assume that the initial PMF 
is statistically
homogeneous, isotropic and random.  
For such a magnetic field, the power spectrum can be taken as a
power-law $P(k) \propto k^{n_B} $ \cite{Mack:2001gc,Kahniashvili:2006hy} where $n_B$ is the spectral index which can be either negative or positive depending upon the
physical processes of the field creation.
From ref.~\cite{Mack:2001gc}, a  two-point correlation function for the PMF can be  defined by
\begin{eqnarray}
\left\langle B^{i}(\mbi{k}) {B^{j}}^*(\mbi{k}')\right\rangle 
	&=&	\frac{(2\pi)^{n_B+8}}{2k_\lambda^{n+3}}
		\frac{B^2_{\lambda}}{\Gamma\left(\frac{n_B+3}{2}\right)}
		k^{n_B}P^{ij}(k)\delta(\mbi{k}-\mbi{k}'), 
		\ \ k < k_C,
		\label{two_point1} 
\end{eqnarray}
where
\begin{eqnarray}
P^{ij}(k)&=&
	\delta^{ij}-\frac{k{}^{i}k{}^{j}}{k{}^2}~~.
	\label{project_tensor}
\end{eqnarray}
 Here, $B_\lambda$ is the comoving mean magnetic-field amplitude obtained by smoothing over a Gaussian sphere of comoving radius $\lambda$,
and $k_\lambda \equiv 2\pi/\lambda$ (with $\lambda=1$ Mpc in this paper).
Hereafter, we work in k-space and denote all quantities by their Fourier transform convention
\begin{eqnarray}
F(\mbi{k})=\int d^3 x \exp (i\mbi{k} \cdot \mbi{x})F(\mbi{x}). 
\label{eq_Fourier}
\end{eqnarray}
The cutoff wave number $k_C$ in the magnetic power
 spectrum is defined by \cite{Subramanian:1997gi,Banerjee:2004df},
\begin{eqnarray}
k_C^{-5-n_B}(\tau)=
\left\{
		\begin{array}{rl}
			\frac{B^2_\lambda k_\lambda^{-n_B-3}}{4\pi(\rho+p)}
			\int^{\tau}_{0}d\tau' 
			\frac{l_{\gamma}}{a},
			& \tau < \tau_\mathrm{dec} \\
			k_C^{-5-n_B}(\tau_\mathrm{dec}), & \tau > \tau_\mathrm{dec}~~,
		\end{array}
\right.
	\label{eq:CutOff_F}
\end{eqnarray}
where $l_\gamma$ is the photon mean free path, and $\tau_\mathrm{dec}$ is the conformal time at the epoch of photon-baryon decoupling. 

We obtain power spectra for the  PMF energy density and the Lorentz force for the scalar mode, respectively,  as follows
\begin{eqnarray}
|E_{\mathrm{[EM:S]}}(\mbi{k},\tau)|^2\delta(\mbi{k}-\mbi{k}')
=
\frac{1}{(2\pi)^3}
\left\langle
	T(\mbi{k},\tau)_{\mathrm{[EM:S1]}}T^*(\mbi{k}',\tau)_{\mathrm{[EM:S1]}}
\right\rangle ~~,
\label{ED_Souce}
\end{eqnarray}
and
\begin{eqnarray}
|\Pi_{\mathrm{[EM:S]}}(\mbi{k},\tau)|^2\delta(\mbi{k}-\mbi{k}')
=&&\frac{1}{(2\pi)^3}
	\left\langle
	\left(
		T(\mbi{k},\tau)_{\mathrm{[EM:S1]}}
		-T(\mbi{k},\tau)_{\mathrm{[EM:S2]}}
	\right)
	\right.
\nonumber\\
&&\times
	\left.
	\left(
		T^*(\mbi{k}',\tau)_{\mathrm{[EM:S1]}}
		-T^*(\mbi{k}',\tau)_{\mathrm{[EM:S2]}}
	\right)
	\right\rangle ~,
\nonumber\\
\label{LF_Souce}
\end{eqnarray}
where S1 and S2 in the subscripts of the energy-momentum tensor denote  the PMF energy density and pressure.
An explicit expression can be obtained for the ensemble averages which are used to evaluate
the above spectra. 
In the case of a power law stochastic magnetic field we have\cite{Koh:2000qw,Yamazaki:2006mi,Brown:2005kr,Kahniashvili:2006hy},
\begin{eqnarray}
\lefteqn{
	\langle
		T(\mbi{k},\tau)_{[\mathrm{EM:S1}]}
		T^*(\mbi{p},\tau)_{[\mathrm{EM:S1}]}
	\rangle
=	
	\frac{1}{2^4(2\pi)^8 a^{8}}
	\left\{
		\frac{(2\pi)^{n+8}}{2k_\lambda^{n+3}}
		\frac{B^2_{\lambda}}{\Gamma\left(\frac{n+3}{2}\right)}
	\right\}^2
}\hspace{1cm}\nonumber\\
&&\times
	\int d^3k'
	k'{}^n|\mbi{k}-\mbi{k}'|^n
	\left\{
		1+\frac{\{\mbi{k}'\cdot(\mbi{k}-\mbi{k}')\}^2}
			   {k'{}^2|\mbi{k}-\mbi{k}'|^2}
	\right\}
	\delta(\mbi{k}-\mbi{p})~.
\nonumber\\
\label{eq:S1S1_fnck}
\end{eqnarray}
The two-point correlation function for the Lorentz force is given by
\begin{eqnarray}
\lefteqn{
	\langle
		T(\mbi{k},\tau)_{[\mathrm{EM:S1}]}
		T(\mbi{p},\tau)^*_{[\mathrm{EM:S1}]}
	\rangle
=
\frac{1}{2^3(2\pi)^7 a^{8}}
\left\{\frac{(2\pi)^{n+8}}{2k_\lambda^{n+3}}\frac{B^2_{\lambda}}{\Gamma\left(\frac{n+3}{2}\right)}\right\}^2
}\hspace{1cm}\nonumber\\
&&\times
\int dk'k'{}^{n+2}\int^{1}_{-1} d\mathcal{C}
|\mbi{k}-\mbi{k}'|^{n-2}
\left\{
(1+\mathcal{C}^2)k^2-4kk'\mathcal{C}+2k'{}^2
\right\}
\delta(\mbi{k}-\mbi{p})~~.
\nonumber\\  
&& 
\label{eq:S1S1_fncC}
\end{eqnarray}
Here, we define $\mathcal{C}$ as
\begin{eqnarray}
	\mathcal{C}
		= \cos{c}
		= \hat{\mbi{k}}\cdot\hat{\mbi{k}}'
		= \frac{\mbi{k}'\cdot\mbi{k}}{k'k}~.
		\label{define_C} 
\end{eqnarray}
Almost all previous works have set the terms which include
$\mathcal{C}$ in the middle parenthesis to unity. 
In this paper, however, we evaluate Eq.~(\ref{eq:S1S1_fncC}) explicitly  using 
integration by parts.  In this way we obtain the following equation.
\begin{eqnarray}
\lefteqn{
	\langle
		T(\mbi{k},\tau)_{[\mathrm{EM:S1}]}
		T^*(\mbi{k},\tau)_{[\mathrm{EM:S1}]}
	\rangle
	=
	\frac{1}{8\pi a^{8}}
	\left\{
		\frac{(2\pi)^{n_B+5}}{2k_\lambda^{n_B+3}}
		\frac{B^2_{\lambda}}{\Gamma\left(\frac{n_B+3}{2}\right)}
	\right\}^2
}\hspace{1cm}
 \nonumber \\
&&\times	
	\int dk'k'{}^{n_B+2}
	\left[
	\frac{n_B^2+4n_B+1}{kk'n_B(n_B+2)(n_B+4)}
	\left\{
		(k+k')^{n_B+2}
		-|k-k'|^{n_B+2}
	\right\}
	\right.
\nonumber\\
&&-
	\frac{1}{k'{}^2n_B(n_B+4)}
	\left\{
		|k-k'|^{n_B+2}
		+|k+k'|^{n_B+2}
\right\}
\nonumber\\
&&+
	\left.
	\frac{k}{k'{}^3n_B(n_B+2)(n_B+4)}
	\left\{
		(k+k')^{n_B+2}
		-|k-k'|^{n_B+2}
	\right\}
	\right].
\label{eq:T1T1}
\end{eqnarray}
A similar derivation leads to the power spectrum of the PMF tension as well as  
the power spectrum of the correlation between pressure and tension
 as follows, 
\begin{eqnarray}
\lefteqn{
	\left\langle 
		T_{[\mathrm{EM:S2}]}(\mbi{k})
		T^*_{[\mathrm{EM:S2}]}(\mbi{k})
	\right\rangle = 
	\frac{1}{2\pi a^{8}}
	\left\{
		\frac{(2\pi)^{n_B+5}}{2k_\lambda^{n_B+3}}
		\frac{B^2_{\lambda}}{\Gamma\left(\frac{n_B+3}{2}\right)}
	\right\}^2
}\hspace{1cm}
	\nonumber\\ 
&&\times
	\int dk'
	k'{}^{n_B+4}
	\frac{4}{(kk')^3n_B(n_B+2)(n_B+4)}
	\left[\frac{}{}
		\left\{
			(k+k')^{n_B+4}
			-|k-k'|^{n_B+4}
		\right\}
	\right.
\nonumber\\
&&-
		\frac{3}{(kk')(n_B+6)}
		\left\{
			|k-k'|^{n_B+6}
			+(k+k')^{n_B+6}
		\right\}
\nonumber\\
&&+
	\left.
		\frac{3}{(kk')^2(n_B+6)(n_B+8)}
		\left\{
			(k+k')^{n_B+8}
			-|k-k'|^{n_B+8}
		\right\}
	\right]~,
\label{eq:T2T2} 
\end{eqnarray}
and
\begin{eqnarray}
\lefteqn{
	\langle
		T_{[\mathrm{EM:S1}]}(\mbi{k})
		T^*_{[\mathrm{EM:S2}]}(\mbi{k})
	\rangle
	+
	\langle
		T_{[\mathrm{EM:S2}]}(\mbi{k})
		T^*_{[\mathrm{EM:S1}]}(\mbi{k})
	\rangle
}\hspace{1cm}
\nonumber\\
&=&\frac{1}{2\pi a^{8}}
	\left\{
		\frac{(2\pi)^{n_B+5}}{2k_\lambda^{n_B+3}}
		\frac{B^2_{\lambda}}{\Gamma\left(\frac{n_B+3}{2}\right)}
	\right\}^2
\nonumber\\
&&\times
    \int dk'k'{}^{n_B+3}
\left[	
    \frac{1}{(kk')^2n_B(n_B+2)}
	\left\{
		(k+k')^{n_B+3}
	   -|k-k'|^{n_B+3}
	\right\}
\right.
\nonumber\\ 
&& 	-
	\frac{3}{k^2k'{}^3n_B(n_B+2)(n_B+4)}
	\left\{
		|k-k'|^{n_B+4}
		+(k+k')^{n_B+4}
	\right\}
\nonumber\\ 
&& 	-
	\frac{1}{k^3k'{}^2n_B(n_B+2)(n_B+4)}
	\left\{
		(k+k')^{n_B+4}
	   -|k-k'|^{n_B+4}
	\right\}
\nonumber\\ 
&&
\left.
	+\frac{3}{k^3k'{}^4n_B(n_B+2)(n_B+4)(n_B+6)}
	\left\{
		(k+k')^{n_B+6}
	   -|k-k'|^{n_B+6}
	\right\}
\right]~~.
\nonumber\\
\label{eq:T1T2}
\end{eqnarray}

For this article 
we have constructed a numerical program, "PriME: Program for primordial Magnetic Effects", with which we can evaluate the PMF source power spectrum using the  numerical method described in 
Refs.~\cite{Yamazaki:2006mi,Yamazaki:2007oc,Yamazaki:2007mm}.
Using this, we can quantitatively evaluate the time
evolution of the cut off scale and thereby reliably calculate the effects of the PMF.

\section{Evolution Equations}
We now summarize the essential evolution equations for each mode.

For the scalar mode we obtain the following equations in $k$-space
\cite{Padmanabhan:1993booka,
Ma:1995ey,
Hu:1997hp,
Hu:1997mn,
Dodelson:2003booka,
Giovannini:2006kc}:
\begin{eqnarray}
k^2\phi + 3H(\dot{\phi}+H\psi) &=& 4\pi G{a^2}
\left\{
	E_\mathrm{[EM:S]}(\mbi{k},\tau)-\delta\rho_\mathrm{tot}
\right\}\\
k^2(\phi-\psi) &=& 
-12\pi G{a^2}
\left\{
	Z_\mathrm{[EM:S]}(\mbi{k},\tau)
	-(\rho_\nu+P_\nu)\sigma_\nu
\right\}
\nonumber\\
 &=&
 -12\pi G{a^2}
 \left\{
 	\frac{1}{3}E_\mathrm{[EM:S]} 
 	(\mbi{k},\tau)+\Pi_\mathrm{[EM:S]}(\mbi{k},\tau)
	-(\rho_\nu+P_\nu)\sigma_\nu
 \right\}~~,
\end{eqnarray}
\begin{eqnarray}
\dot{\delta}^\mathrm{(S)}&
			=&-(1+w)\left(v^\mathrm{(S)}+3\dot{\phi}\right)
			-3H\left(\frac{\delta p}{\delta\rho}-w\right)\delta^\mathrm{(S)}
\nonumber\\
&&			-\frac{3}{8\pi \rho}
			\left\{
				\dot{E}_{\mathrm{[EM:S]}}(\mbi{k},\tau)
				+6HE_{\mathrm{[EM:S]}}(\mbi{k},\tau)
			\right\}~,
			\label{eq:scalar_contiunuity1}\\
\dot{v}^\mathrm{(S)}&=&-H(1-3w)v^\mathrm{(S)}
			-\frac{\dot{w}}{1+w}v^\mathrm{(S)}
			+\frac{\delta p}{\delta\rho}
			\frac{k^2\delta^\mathrm{(S)}}{1+w}
			-k^2\sigma+k^2\psi
\nonumber\\
&&			
			+k^2 \frac{\Pi_{\mathrm{[EM:S]}}(\mbi{k},\tau)}
					 {4\pi \rho} ~,
\label{eq:scalar_motion}
\end{eqnarray}
where 
$
Z_\mathrm{[EM:S]}\equiv
 \sigma_\mathrm{B}(\rho_\gamma+P_\gamma)
$ and 
$w \equiv p/\rho$ is the usual equation of state parameter.
Note that for the photon 
$\delta^\mathrm{(S)}_\gamma = 4\Theta^\mathrm{(S)}_0$, and 
$v^\mathrm{(S)}_\gamma=k\Theta^\mathrm{(S)}_1$.
Massless neutrinos obey Eqs.~(\ref{eq:scalar_contiunuity1}) and (\ref{eq:scalar_motion}) as written without the Thomson coupling term.
In the continuity and Euler relations (Eqs.~\ref{eq:scalar_contiunuity1} and \ref{eq:scalar_motion}) for the scalar mode, we can
just add the energy density and pressure of the PMF to the 
energy density and pressure of the cosmic fluids. Since the baryon
fluid behaves like a nonrelativistic fluid during the epoch of interest, we
can neglect $w$ and $\delta P^\mathrm{(S)}_b/\delta\rho^\mathrm{(S)}_b$, except for the acoustic term 
$c_sk^2\delta^\mathrm{(S)}_b$. Also, the shear stress of the baryons is negligible \cite{Ma:1995ey}. 
Since we concentrate on scalar type
perturbations in this paper, we do not consider the magneto-rotational
instability from the shear stress of the PMF and baryon fluid
\cite{Chandrasekhar:1961ab}.

 By considering the Compton interaction
 between baryons and photons  
in equations (\ref{eq:scalar_contiunuity1}) and
 (\ref{eq:scalar_motion}) we obtain the same form for  the evolution
 equations of photons and baryons as in previous
 work \cite{Padmanabhan:1993booka,Ma:1995ey,Hu:1997hp,Hu:1997mn,Dodelson:2003booka}.
  \begin{eqnarray}
k^2\phi + 3H(\dot{\phi}+H\psi) &=& 4\pi G{a^2}
\left\{
	E_\mathrm{[EM:S]}(\mbi{k},\tau)-\delta\rho_\mathrm{tot}
\right\} \label{eq:phi}\\
k^2(\phi-\psi) &=& 
-12\pi G{a^2}
\left\{
	Z_\mathrm{[EM:S]}(\mbi{k},\tau)
	-(\rho_\nu+P_\nu)\sigma_\nu
\right\}
\nonumber\\
 &=&
 -12\pi G{a^2}
 \left\{
 	\frac{1}{3}E_\mathrm{[EM:S]} 
 	(\mbi{k},\tau)+\Pi_\mathrm{[EM:S]}(\mbi{k},\tau)
	-(\rho_\nu+P_\nu)\sigma_\nu
 \right\}~~,
 \label{eq:phi_psi}\\
\dot{\delta}^\mathrm{(S)}_\mathrm{CDM}
 	&=&
 		-v^\mathrm{(S)}_\mathrm{CDM}+3\dot{\phi}~,\label{eq:CDM_rho}\\
\dot{v}^\mathrm{(S)}_\mathrm{CDM}
 	&=&
        -\frac{\dot{a}}{a}v^\mathrm{(S)}_\mathrm{CDM}+k^2\psi~,\label{eq:CDM_v}\\
\dot{\delta}^\mathrm{(S)}_{\gamma}
 	&=&
 		-\frac{4}{3}v^\mathrm{(S)}_{\gamma}
 		+4\dot{\phi}~,\label{eq:photon_rho}\\
\dot{\delta}^\mathrm{(S)}_{\nu}
 	&=&
 		-\frac{4}{3}v^\mathrm{(S)}_{\nu}
 		+4\dot{\phi}~,\label{eq:photon_nu}\\
\dot{v}^\mathrm{(S)}_{\gamma}
	&=&
		k^2\left(\frac{1}{4}\delta^\mathrm{(S)}_{\gamma}-\sigma_{\gamma}\right)
		+an_e\sigma_T(v^\mathrm{(S)}_\mathrm{b}-v^\mathrm{(S)}_{\gamma})~+k^2\psi,
		\label{eq:photon_v} \\
\dot{v}^\mathrm{(S)}_{\nu}
	&=&
		k^2\left(\frac{1}{4}\delta^\mathrm{(S)}_{\nu}-\sigma_{\nu}\right)+k^2\psi,\label{eq:photon_nu} \\
\dot{\delta}^\mathrm{(S)}_\mathrm{b}
	&=&
		-v^\mathrm{(S)}_\mathrm{b}+3\dot{\phi}~,
 			\label{eq:baryon_rho}  \\
\dot{v}^\mathrm{(S)}_\mathrm{b}
	&=&
			-\frac{\dot{a}}{a}v^\mathrm{(S)}_\mathrm{b}
 			+c^2_sk^2\delta^\mathrm{(S)}_\mathrm{b}
 			+\frac{4\bar{\rho}_\gamma}{3\bar{\rho}_\mathrm{b}}
 			an_e\sigma_T(v^\mathrm{(S)}_{\gamma}-v^\mathrm{(S)}_\mathrm{b})+k^2\psi
\nonumber\\
&&			
			+\frac{3}{4}k^2\frac{\Pi_{\mathrm{[EM:S]}}(\mbi{k},\tau)}{R\rho_\gamma}~,
			\label{eq:baryon_v}
\end{eqnarray} 
where $R\equiv(3/4)(\rho_b/\rho_\gamma)$ is the inertial density ratio of baryons to photons, $n_e$ is the free electron density, $\sigma_T$ is the Thomson
scattering cross section, and $\sigma_{\gamma}$ of the second term on
the right hand side of equation (\ref{eq:photon_v}) is the
shear stress of the photons with the PMF. 
Since $n_\mathrm{B}\lesssim 0$ is favored by constraints from the gravitational wave 
background \cite{Caprini:2001nb}, and the  PMF effects are not influenced by the time evolution of
the cut off scale $k_C$ for this range of $n_\mathrm{B}$,
we can approximately set $E_\mathrm{[EM:S]}\propto a^{-4}$ in the following analysis.
\subsection{Initial Conditions}
We need to specify the initial perturbations for solving the evolution equations presented in the previous section.
We start the solution at early times when the $k$ modes of interest
are still outside the horizon, i.e.~the dimensionless parameter  $k\tau \ll 1$.
We consider only the radiation-dominated epoch since the numerical integration for all of the $k$ modes of interest will start within this era. 
Baryons and photons are tightly coupled at this early time and 
the expansion rate is $H = \tau^{-1}$.
We derive initial conditions for all of the modes utilizing  the method of
Refs.~\cite{Padmanabhan:1993booka,
Ma:1995ey,
Hu:1997hp,
Hu:1997mn,
Dodelson:2003booka,
Giovannini:2006kc}.
We can assume that all density fields are  zero  initially, since
the PMF only affects the velocity field of ionized baryons,   via the  Lorentz force, and the density fields are not directly affected by the PMF.
In the radiation dominated epoch, photons and neutrinos are important in the energy-momentum tensor. The evolution equations for the photons and neutrinos are
\begin{eqnarray}
\dot{\delta}^\mathrm{(S)}_{\gamma}
 	&=&
 		-\frac{4}{3}v^\mathrm{(S)}_{\gamma}
 		+4\dot{\phi}~,\label{eq:photon_rho_early}\\
\dot{\delta}^\mathrm{(S)}_{\nu}
 	&=&
 		-\frac{4}{3}v^\mathrm{(S)}_{\nu}
 		+4\dot{\phi}~,\label{eq:nu_rho_early}\\
\dot{v}^\mathrm{(S)}_{\gamma}
	&=&
		k^2\frac{1}{4}\delta^\mathrm{(S)}_{\gamma}
		+k^2\psi,
		\label{eq:photon_v_early} \\
\dot{v}^\mathrm{(S)}_{\nu}
	&=&
		k^2
		\left(
				 \frac{1}{4}\delta^\mathrm{(S)}_{\nu}
				-\sigma^\mathrm{(S)}_{\nu}
				+k^2\psi
		\right),
		\label{eq:nu_v_early}\\
\sigma^\mathrm{(S)}_{\nu}
	&=&
		\frac{4}{15}v^\mathrm{(S)}_\gamma
.\label{eq:nu_v_early}
\end{eqnarray}
Here, we have omitted higher multipole moments $\ell > 1$ for photons and $\ell > 2$ for neutrinos. 
At  lowest order in $k\tau$, the initial conditions for  Eqs.(\ref{eq:phi}-\ref{eq:baryon_v}) are
\begin{eqnarray}
\delta^\mathrm{(S)}_\gamma  = 
\delta^\mathrm{(S)}_\nu  =
\frac{4}{3}\delta^\mathrm{(S)}_b  =
\frac{4}{3}\delta^\mathrm{(S)}_\mathrm{CDM}  &=&
	R_\gamma R_\mathrm{B}
	+4R_\gamma
		\frac{4\sigma_\mathrm{B}+R_\nu R_\mathrm{B}}
			 {4R_\nu+15},\\
v^\mathrm{(S)}_\gamma = 
v^\mathrm{(S)}_b = 
v^\mathrm{(S)}_\mathrm{CDM} &=& 
-\frac{19}{4}
		\frac{4\sigma_\mathrm{B}+R_\nu R_\mathrm{B}}
			 {4R_\nu+15}
			 k^2\tau,\\
v^\mathrm{(S)}_\nu &=& 
-\frac{15}{4}
 \frac{R_\gamma}{R_\nu}
		\frac{4\sigma_\mathrm{B}+R_\nu R_\mathrm{B}}
			 {4R_\nu+15}
			 k^2\tau,\\
\sigma^\mathrm{(S)}_\nu &=& 
 -\frac{R_\gamma}{R_\nu}\sigma_\mathrm{B}
 +
 \frac{R_\gamma}{2R_\nu}
		\frac{4\sigma_\mathrm{B}+R_\nu R_\mathrm{B}}
			 {4R_\nu+15}
			 k^2\tau^2,\\
\psi = 
-2\phi &=& 
 -2R_\gamma
 \frac{R_\gamma}{2R_\nu}
		\frac{4\sigma_\mathrm{B}+R_\nu a R_\mathrm{B}}
			 {4R_\nu+15}~,
\end{eqnarray}
where 
\begin{eqnarray}
R_\gamma
	&\equiv&
		\frac{\rho_\gamma}{\rho_\gamma+\rho_\nu},\nonumber\\
R_\nu
	&\equiv&
		\frac{\rho_\nu}{\rho_\gamma+\rho_\nu},\nonumber\\
R_\mathrm{B}
	&\equiv&
		\frac{E_\mathrm{[EM:S]}}{\rho_\gamma}~.\nonumber
\end{eqnarray}
\section{\label{s:transfer_function}Matter power spectrum}
Possible origins of the PMF have been studied by many authors, however,
there is no consensus yet as to  the origin of the PMF.
Thus, we cannot know how the PMF correlates with the primordial density fluctuations.
However, almost all previous works investigated the effects of a PMF on
density perturbations under  the assumption that there is no correlation
between the PMF and the primordial density fluctuations \cite{Tashiro:2005ua}. 
However, in order to study the PMF effects in a more general manner, 
 we introduce a parameter "$s$" which characterizes the correlation between the PMF and the primordial density fluctuations\cite{Yamazaki:2006mi,Giovannini:2006kc}. 
In the linear approximation, the power spectra of the baryon ($P_\mathrm{b}(k)$) and CDM ($P_\mathrm{CDM}(k)$) density fluctuations in the presence of a PMF are then written,
\begin{eqnarray}
P_\mathrm{b}(k)&=&
	\left\langle 
		\delta_\mathrm{[b:FL]}(k)
		\delta_\mathrm{[b:FL]}^*(k)
	\right\rangle
	+
	\left\langle 
		\delta_\mathrm{[b:PMF]}(k)
		\delta_\mathrm{[b:PMF]}^*(k)
	\right\rangle 
	 \nonumber\\
	&+&
		2\left\langle 
		\delta_\mathrm{[b:FL]}(k)
		\delta_\mathrm{[b:PMF]}^*(k)
		\right\rangle ,
	 \label{CTFb}\\
P_\mathrm{CDM}(k)&=&
	\left\langle 
		\delta_\mathrm{[CDM:FL]}(k)
		\delta_\mathrm{[CDM:FL]}^*(k)
	\right\rangle
	+
	\left\langle 
		\delta_\mathrm{[CDM:PMF]}(k)
		\delta_\mathrm{[CDM:PMF]}^*(k)
	\right\rangle  \nonumber\\ 
	&+&
		2\left\langle 
		\delta_\mathrm{[CDM:FL]}(k)
		\delta_\mathrm{[CDM:PMF]}^*(k)
		\right\rangle ,
 \label{CTFCDM}
\end{eqnarray}
where we normalize the cross correlation terms with the parameter $s$,
\begin{eqnarray}
	\left\langle 
	\delta_\mathrm{[b:FL]}(k)
	\delta_\mathrm{[b:PMF]}^*(k)
	\right\rangle &\equiv& s
	\sqrt{
	        \left\langle 
		\delta_\mathrm{[b:FL]}(k)
		\delta_\mathrm{[b:FL]}^*(k)
	        \right\rangle
	        \left\langle 
		\delta_\mathrm{[b:PMF]}(k)
		\delta_\mathrm{[b:PMF]}^*(k)
	        \right\rangle
		},
\label{crossb}\\
	\left\langle 
	\delta_\mathrm{[CDM:FL]}(k)
	\delta_\mathrm{[CDM:PMF]}^*(k)
	\right\rangle 
	&\equiv& s 
	\sqrt{
	        \left\langle 
		\delta_\mathrm{[CDM:FL]}(k)
		\delta_\mathrm{[CDM:FL]}^*(k)
	        \right\rangle
	        \left\langle 
		\delta_\mathrm{[CDM:PMF]}(k)
		\delta_\mathrm{[CDM:PMF]}^*(k)
	        \right\rangle
		}~. \nonumber  \\
\label{crossCDM}
\end{eqnarray}
Here,  $\delta_\alpha$, $\alpha \in (\mathrm{[b:FL]}, \mathrm{[CDM:FL]})$
designate the baryon and CDM density fluctuations without the PMF respectively.  Similarly,  
$\delta_\beta$, $\beta \in (\mathrm{[b:PMF]}, \mathrm{[CDM:PMF]})$
denote the baryon and CDM density fluctuations with the PMF included.  
When $0<s\le 1$, $s=0$, or $-1\le s<0$ in eqs.(\ref{crossb}) and (\ref{crossCDM}), one has positive, vanishing, or negative correlations, respectively.

The square root of the power spectrum functions for the Lorentz force 
$\Pi_{\mathrm{[EM:S]}}(\mbi{k},\tau)$
in Eq.(\ref{eq:baryon_v}) does not specify the sign.
In other words, there is no information as to whether  the magnetic pressure or the tension is dominant, and whether the directions of forces from them are the same or different.
However,  such information  should be taken into account.

The Lorentz force term in Eq.(\ref{eq:baryon_v}) can be divided into two terms,
 the magnetic pressure and the tension.  Their amplitudes are given by Eqs.~(\ref{eq:T1T1}) and (\ref{eq:T2T2}), respectively.
By comparing those equations, one can decide 
which of them is dominant in the Lorentz force term.
We find that the former dominates when $n < -1.5$,
while the latter dominates for $n > -1.5$. 
However, one cannot determine the relative signs of those two terms unless one also specifies a model for the generation of the PMF \cite{Yamazaki:2006mi}.
Thus, we must decompose the factors into various combinations, i.e. 
\begin{eqnarray}
s=s_\mathrm{[LF]}\times s_\mathrm{[DF]},
\end{eqnarray}
where
\begin{eqnarray}
s_\mathrm{[LF]}=
\left\{
		\begin{array}{rl}
			-1, & n < -1.5\ ~\mathrm{(I)},\\
			-1, & n > -1.5\ ~\mathrm{(II)}, \\
			1, & n > -1.5\ \   ~\mathrm{(III)},
		\end{array}
\right.
	\label{eq:SLF}
\end{eqnarray}
and
\begin{eqnarray}
		\begin{array}{rlccc}
			0 & < & s_\mathrm{[DF]} &\le& 1\ ~\mathrm{(i)},\\
			  &   & s_\mathrm{[DF]} & = & 0\ ~\mathrm{(ii)},\\
			-1&\le& s_\mathrm{[DF]} & < & 0\ ~\mathrm{(iii)}.
		\end{array}\nonumber
\end{eqnarray}
In the different regimes,  $s_\mathrm{[LF]}$ represents either:
(I) the pressure dominated case;
(II) the tension dominated case, where the  magnetic field pressure and tension forces act in the same direction; or 
(III) the tension dominated case, where the magnetic field pressure and tension forces act in the opposite direction.
On the other hand, $s_\mathrm{[DF]}$ represents either:
(i) a positive correlation between  
the matter and PMF distributions;  (ii) no correlation;
or (iii) a negative correlation.
Thus, if $s<0$, 
the  matter and PMF distributions could  be correlated positively ($s_\mathrm{[DF]}>0$) 
and  the PMF pressure dominates in the Lorentz term (for $n < -1.5$).  
Another possibility is that 
the matter and PMF distributions negatively correlate ($s_\mathrm{[DF]}<0$) 
and the PMF tension dominates in the Lorentz term (for $n > -1.5$) and 
the tension acts on the density field in the same direction as the magnetic field pressure.
Yet another possibility is that 
the PMF tension dominates in the Lorentz term ($n > -1.5$),
but the tension acts on the density field in the opposite direction from  the pressure force.
In these cases the PMF effects act like a gas pressure to oppose the gravitational collapse and 
cause the density perturbations to more slowly evolve.

On the other hand, 
if $s>0$, 
the matter and PMF distributions could positively correlate ($s_\mathrm{[DF]}>0$) 
and 
the PMF tension dominates in the Lorentz term ($n > -1.5$) while
the tension acts on the density field in the opposite direction from the pressure force.
Alternatively,  
 the matter and PMF distributions could negatively correlate
($s_\mathrm{[DF]}<0$)  
and 
the PMF pressure dominate in the Lorentz term ($n < -1.5$). 
In these cases the Lorentz force from the PMF accelerates the gravitational collapse.
After decoupling, $\delta$ does not
oscillate and the perturbation evolution is straightforward for all of the above cases. 
\section{Results and Discussions}
For completeness, in this section we briefly review the effects of the PMF on the cosmological density field fluctuations (see \cite{Yamazaki:2006mi} for details). 
We will then illustrate that relation between $\sigma_8$ and the PMF parameters.
We will show that the constraints on  $\sigma_8$ from observation give a strong constraint on the PMF parameters. 
Since parameters of the PMF have a strong degeneracy, the existence of such a prior, can be used to effectively constrain the PMF. Also, since $\sigma_8$ is constrained by diverse observational data on linear cosmological scales, we can obtain a reliable prior 
for use in determining likelihood functions for the parameters of the PMF from CMB observations.
The PMF effects dominate the matter power spectra for wavenumbers $k >  0.1$ Mpc$^{-1}$\cite{Yamazaki:2006mi}.  This is because the PMF energy density fluctuations depend only on the scale factor $a$ and can survive below the Silk damping scale. Therefore, the PMF continues to source the fluctuations through the Lorentz force even below the Silk damping scale.
In the case of no correlation between the PMF and the matter density fluctuations, the matter power spectrum is increased by the PMF, independently of whether the PMF pressure or tension dominates.

Here we note the different effects of the PMF on the power spectrum function $P(k)$ and the matter density fluctuation $\delta$. 
While the total density fluctuation $\delta$ can be smaller or larger
depending upon whether the effect of the PMF is dominated by its pressure or tension, the power spectrum function $P(k)$ always increases
when the PMF does not correlate with the primordial density fluctuations.  This is  because $P(k)\propto \delta^2$ and is not affected by the sign of $\delta$.
\subsection{Effects of PMF parameters on $\sigma_8$}
The alternative normalization parameter $\sigma_8$ is the root-mean-square of the matter density fluctuation in a comoving sphere of radius $8h^{-1}$ Mpc.  It is given by a weighted integral of the matter power spectrum \cite{Peebles:1980booka}. 
We can study the physical processes of density field fluctuations on cosmological scales within the linear regime to determine $\sigma_8$.
Recently $\sigma_8$ has been constrained by observations \cite{Cole:2005sx,Tegmark:2006az,Rozo:2007yt,Ross:2008ze} to be in the range  $0.7 < \sigma_8 < 0.9$.
From this we can obtain strong constraint for the PMF parameters by numerically calculating $\sigma_8$ under the influence of PMF effects.

We expect that the discrepancy between theoretical estimates and observational temperature fluctuations of the CMB for higher multipolarity ($\ell > 1000 $) is solved by combining a PMF of strength $2.0~nG <|B_\lambda|< 3.0~nG$ and the SZ effects. In this case, $\sigma_8$ derived by such a field strength for the PMF is $0.77 - 0.88$. This is consistent with our assumed  prior in the range $\sigma_8$ as $0.7 < \sigma_8 < 0.9$.
Since $\sigma_8$ is affected by other cosmological parameters, $\Omega_b$, $\Omega_\mathrm{CDM}$, $n_\mathrm{S}$, and $A_\mathrm{S}$, 
we should consider the degeneracy between the PMF and other cosmological parameters as mentioned above.
Fortunately, these cosmological parameters are constrained by recent CMB observations on larger scales ($\ell < 1000$) \cite{Spergel:2006hy,Hinshaw:2006ia,Page:2006hz}, while it was shown in our previous work \cite{Yamazaki:2006bq,Yamazaki:2006mi,Yamazaki:2007oc} that the effect of the PMF mainly affects  the CMB anisotropies on smaller scales ( $\ell > 1000$).
Hence, we expect that the degeneracy between the PMF parameters and the other 
cosmological parameters is small.
 For this reason in the present analysis we are justified in fixing  the other cosmological parameters at 
their best fit values.

\begin{figure*}[h]
\includegraphics[width=1.0\textwidth]{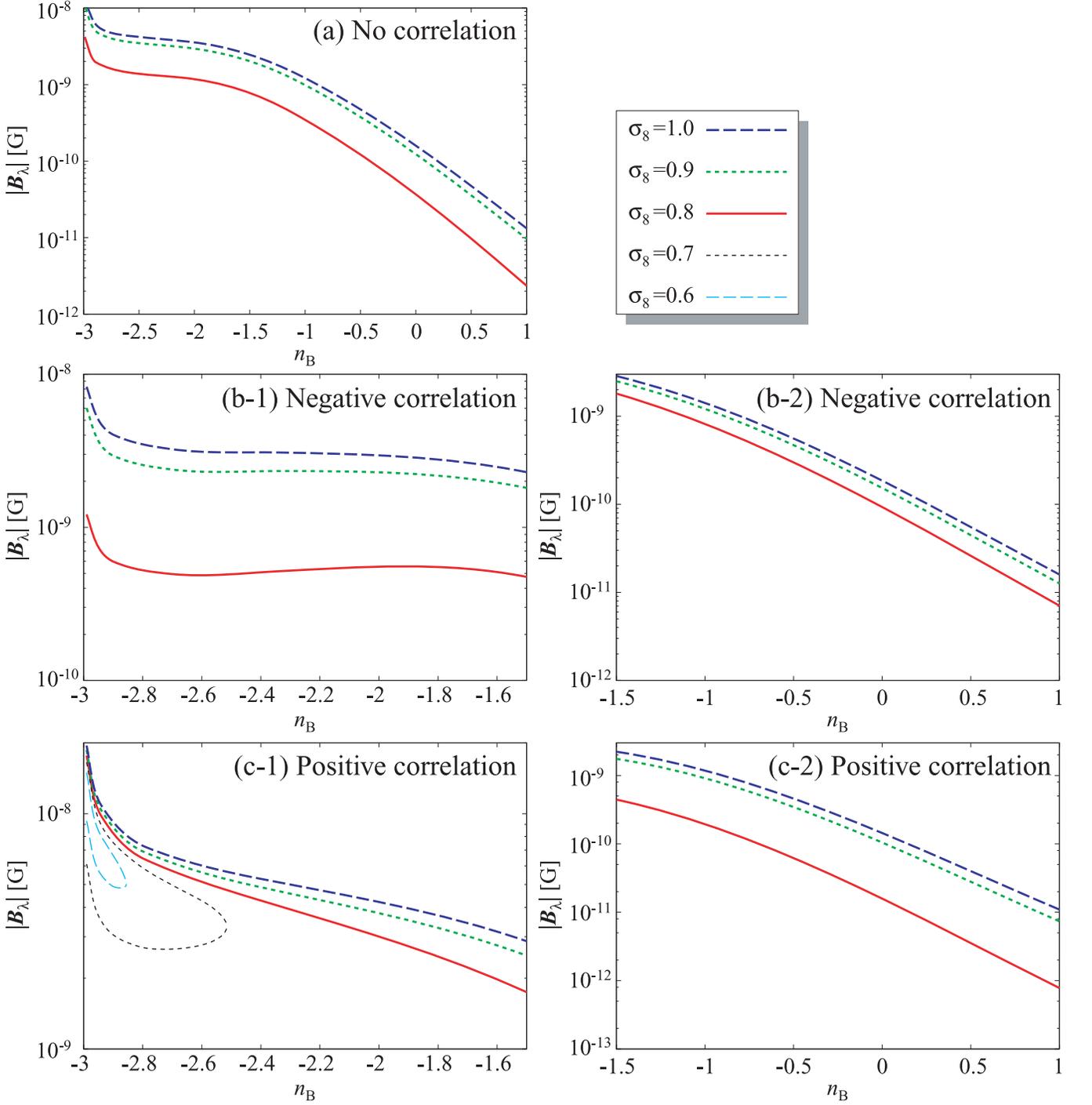}
\caption{\label{fig1} Curves of constant values for  $\sigma_8$ in the parameter plane of PMF amplitude $B_\lambda$ vs.~spectral index $n_\mathrm{B}$.
Blue dashed, green dotted, red bold, black dotted thin, and azure dashed thin curves show constant values of $\sigma_8 = $1.0, 0.9, 0.8, 0.7 and 0.6, respectively.}
\end{figure*}

Figure \ref{fig1} shows the behavior of the  PMF parameters $B_\lambda$ and $n_\mathrm{B}$  for various constant  values of $\sigma_8$ as labeled.
Since the PMF power spectrum depends upon $n_\mathrm{B}$ from Eqs.~(\ref{eq:T1T1}-\ref{eq:T1T2})\cite{Mack:2001gc},  for $n_\mathrm{B} > -1.5$, the PMF effects on density fluctuations for the small scales decrease with lower values for $n_\mathrm{B}$.
While, for $n_\mathrm{B} < -1.5$, $P(k)_\mathrm{PMF} \propto B_\lambda^{{[(2n_\mathrm{B}+3)}/{(n_\mathrm{B}+5})]}$ from
Eqs.~(\ref{eq:T1T1}-\ref{eq:T1T2})\cite{Mack:2001gc}.
Therefore, for a spectral index near $n_\mathrm{B} = -3.0$, the matter power spectrum in the presence of a  PMF is smaller, and larger amplitudes of $B_\lambda$ are allowed.
Since the correlation effects of the PMF for negative and positive correlations change at $n_\mathrm{B} = -1.5$ (see Section \ref{s:transfer_function}), we divide the discussion below of each correlation into two parts based upon whether the spectral index is greater or less than  $n_\mathrm{B} = -1.5$.

\subsubsection{No Correlation}
When there is no correlation between the PMF and the density fluctuations from primary perturbations, the third terms in both Eqs.~(\ref{crossb}) and (\ref{crossCDM}) vanish and the PMF only acts to increases the total matter power spectrum. 
In this paper, we adopt the constraint  that PMF parameters giving $\sigma_8 > 1$ are excluded  by observations.
Panel (a) of Figure.\ref{fig1} shows that a PMF amplitude of $B_\lambda\ ^>_\sim 1$ nG  is excluded when $n_\mathrm{B} > -0.9$.   Furthermore, PMF amplitudes of $B_\lambda\ ^>_\sim 0.11$ nG are excluded when $n_\mathrm{B} > 0.2$.
The magnetic field strength in galaxy clusters is $\sim 1 \mu$G.  Therefore, if  isotropic collapse is the only process which amplifies the magnetic field strength, the lower limit to the PMF is  $~\sim 1$nG at $z \sim 0$. 
Hence, we can obtain a strong constraint on this PMF evolution model for a PMF spectral index in the  range $n_\mathrm{B} < -0.9$.
\subsubsection{Negative Correlation}
For the case of negative correlations,  the pressure of the PMF dominates for $n_\mathrm{B}< -1.5$, and the PMF causes an increase in the  density fluctuations. 
For $n_\mathrm{B}> -1.5$, however, the tension of the PMF dominates, and  the PMF causes a decrease in the  density fluctuations.
These behaviors can be traced to the third terms in each Eqs.(\ref{crossb}) and (\ref{crossCDM}).
Using the allowed range of PMF parameters as mentioned above,
a PMF of $B_\lambda\ ^>_\sim 1$ nG is excluded for  $n_\mathrm{B} > -0.81$, 
and a PMF of $B_\lambda\ ^>_\sim 0.11$ nG is excluded for  $n_\mathrm{B} > 0.26$.
\subsubsection{Positive Correlation}
When there is a positive correlation, the pressure of the PMF dominates for $n_\mathrm{B}< -1.5$, and the PMF leads to a decrease in the  density fluctuations. 
On the other hand,  the tension of the PMF dominates for $n_\mathrm{B}> -1.5$. 
In this case  the PMF causes an increase in the  density fluctuations.
We can attribute  these behaviors again to the  third terms in Eqs.~(\ref{crossb}) and (\ref{crossCDM}).
Using the allowed range of PMF parameters as noted above,
a PMF of  $B_\lambda\ ^>_\sim 1$ nG is excluded for   $n_\mathrm{B} > -0.94$, 
and a PMF of  $B_\lambda\ ^>_\sim 0.11$ nG is excluded for   $n_\mathrm{B} > 0.13$

For both negative correlations with  $n_\mathrm{B} > -1.5$ and positive correlations with  $n_\mathrm{B} < -1.5$, the PMF decreases the total matter power spectrum until the strength of the PMF effect is comparable to the  primary matter power spectrum.
Beyond this point, the density fluctuations from the PMF exceed the density fluctuations from the primary power spectrum and the PMF effect dominates the total matter power spectrum. The amplitude of the total matter power spectra when a PMF is present  is greater than the total matter power spectrum without the presence of a PMF.

In these cases, we obtain different constraints on the strength of the PMF for each value of $n_\mathrm{B}$ from the other models.
We assume that ranges of the PMF parameters giving $\sigma_8 < 0.6 $ are excluded by observations.
Panel (c-1) of Figure.\ref{fig1} shows the excluded range of the PMF parameters within the contour defined by $\sigma_8 = 0.6$.

Our result is consistent with previous constraints on PMF parameters \cite{Caprini:2001nb,Yamazaki:2006ah}, and our new more precise constraints  obtained from the matter power spectrum  are independent of previous methods to constrain the PMF parameters.   Hence, we can now constrain more precisely the  physical processes by which a PMF affects the evolution of structure on cosmological scales.
\section{Summary}
A primordial magnetic field (PMF) affects the  evolution of density field fluctuations in the early universe.
Therefore, we can constrain PMF parameters, e.g. the PMF amplitude $B_\lambda$ and  power spectral index $n_\mathrm{B}$, by comparing a theoretical calculation of the density field fluctuations affected by a  PMF and observational data, e.g. a number density fluctuation of galaxies as indicated by the  $\sigma_8$ parameter.
We have illustrated the relation between $\sigma_8$ and the PMF parameters.  
We have shown that the observed range of $\sigma_8$ gives a
strong limit on the PMF parameters given by  $n_\mathrm{B} < 0.9$
(no correlation case),  $< -0.86$ (negative correlation case), and $< 0.94$
(positive correlation case) for $B_\lambda > 1$nG.
We have also shown how the PMF parameters which were constrained by previous methods are affected by our new constraint from $\sigma_8$.
Since density field fluctuations are the origin of the LSS on cosmological scales,  we can use this as a prior to study the physical processes of the PMF and  to place better constraints on  the PMF parameters.

\begin{acknowledgments}
We acknowledge Drs. K. Umezu, and H. Hanayama for their valuable discussions.
K.I. acknowledges the support by Grants-in-Aid for JSPS Fellows.
This work has been supported in part by Grants-in-Aid for Scientific
Research (17540275, 20244035) of the Ministry of Education, Culture, Sports,
Science and Technology of Japan, and the Mitsubishi Foundation.  This
work is also supported by the JSPS Core-to-Core Program, International
Research Network for Exotic Femto Systems (EFES).  Work at UND supported in part by the US Department of Energy under research grant DE-FG02-95-ER40934.
\end{acknowledgments}
\bibliographystyle{apsrev}
\bibliography{ms_astro_ph}
\end{document}